# Multiharmonic frequency-chirped transducers for surface-acoustic-wave optomechanics

Matthias Weiß[1,2], Andreas L. Hörner[1], Eugenio Zallo[3,‡], Paola Atkinson[3,4], Armando Rastelli[5,3], Oliver G. Schmidt[3], Achim Wixforth[1,2,6], Hubert J. Krenner[1,2,6,*]

[1] Lehrstuhl für Experimentalphysik 1 and Augsburg Centre for Innovative Technologies (ACIT), Universität Augsburg, Universitätsstraße 1 86159 Augsburg, Germany

[2] Nanosystems Initiative Munich (NIM), Schellingstraße 4, 80339 München, Germany

[3] Institute for Integrative Nanosciences, Leibniz IFW Dresden, Helmholtzstraße 20, 01069 Dresden, Germany

[4] Institut des NanoSciences de Paris, Sorbonne Universités, UPMC Univ. Paris 06, CNRS UMR7588, 4 Place Jussieu, F-75005 Paris, France

[5] Institute of Semiconductor and Solid State Physics, Johannes Kepler Universität Linz, Altenbergerstraße 69, 4040 Linz, Austria

[6] Center for Nanoscience (CeNS), Ludwig-Maximilians-Universität München, Geschwister-Scholl-Platz 1, 80539 München, Germany

[*] hubert.krenner@physik.uni-augsburg.de

[‡] present address: Paul-Drude-Institut für Festkörperelektronik, Hausvogteiplatz 5-7, 10117 Berlin, Germany

Abstract:

Wide passband interdigital transducers are employed to establish a stable phase-lock between a train of laser pulses emitted by a mode-locked laser and a surface acoustic wave generated electrically by the transducer. The transducer design is based on a multi-harmonic split-finger architecture for the excitation of a fundamental surface acoustic wave and a discrete number of its overtones. Simply by introducing a variation of the transducer's periodicity $p$ a frequency chirp is added. This combination results in wide frequency bands for each harmonic. The transducer's conversion efficiency from the electrical to the acoustic domain was characterized optomechanically using single quantum dots acting as nanoscale pressure sensors. The ability to generate surface acoustic waves over a wide band of frequencies enables advanced acousto-optic spectroscopy using mode-locked lasers with fixed repetition rate. Stable phase-locking between the electrically generated acoustic wave and the train of laser pulses was confirmed by performing stroboscopic spectroscopy on a single quantum dot at a frequency of 320 MHz. Finally, the dynamic spectral modulation of the quantum dot was directly monitored in the time domain combining stable phase-locked optical excitation and time-correlated single photon counting. The demonstrated scheme will be particularly useful for the experimental implementation of surface acoustic wave-driven quantum gates of optically addressable qubits or collective quantum states or for multi-component Fourier synthesis of tailored nanomechanical waveforms.

## Main text:

### Introduction

The realization of hybrid quantum systems [1,2] is guided by the vision to exploit the strengths of individual constituents while at the same time bypassing their detrimental shortcomings. In this quest at the forefront of contemporary fundamental and applied research, mechanical systems stand out: vibrational excitations and phonons couple to virtually any type of quantum system. In condensed matter, a wide range of qubit systems have been successfully interfaced with localized and narrow-band vibrational modes [3–7]. For a full-fledge quantum circuitry in which qubits are interconnected via mechanical links, surface acoustic waves (SAW) are of paramount importance. SAWs can be elegantly excited directly on piezoelectric materials using so-called interdigital transducers (IDTs) [8] and are only weakly susceptible to dissipation. The periodicity $p$ of such interdigitating combs of electrodes determines the frequency of the excited SAW $f_{SAW} = c_{SAW}/2p$, with $c_{SAW}$ being the SAW phase velocity in the material along the SAW propagation direction. Key wave phenomena can be directly employed using these surface confined mechanical waves. For instance, the generation of arbitrary dynamic nanomechanical waveforms have been realized very recently by additive Fourier synthesis [9]. Similarly, acoustic Bragg resonators have been realized. Due to the SAW's low phase velocity, typically ranging $c_{SAW} \approx 3000 - 7000$ m/s, high quality SAW-resonators are widely used as small and compact radio frequency filters [10,11], which are indispensable in modern wireless communication networks. This application makes SAWs a *phononic* technology of highest industrial relevance.

Over the past more than 30 years, SAWs also have been advanced towards a versatile probe and manipulation tool in fundamental research. Examples include the investigation of collective quantum effects such as the Quantum Hall Effects [12–14], and acoustic transport of charges [15–20] and spins [21,22] in low-dimensional semiconductor nanosystems. Recently, SAWs have been successfully interfaced with single quantum systems, individual charges [23–25] and spins [26], or superconducting quantum bits [27]. In addition to these electrically addressable quantum systems, optically active systems are of paramount importance as these allow to transduce excitations from the acoustic domain at hypersonic frequencies in the 100 MHz to the low GHz range to the optical domains at frequencies of 100's of THz. In this area, single optically active quantum dots (QDs) and defect centers [28], and photonic [29–32] and optomechanical cavities [33–36] have been in focus. On one side, acoustically regulated carrier injection may be used to deliver a precisely triggered train of single photons [37–39] from discrete states in a QD programmed by the SAW [40–44]. On the other hand, the dynamic mechanical displacement and its resulting strain lead to a spectral modulation of the emitter due to deformation potential coupling [45], even in the resolved sideband regime [46,47]. These recent breakthroughs in the deliberate coupling of single quantum systems to propagating SAWs were put on a solid theoretical foundation of SAW-based quantum acoustics [48] in which SAW phonons take the role of microwave photons. Albeit such quantum acoustics can be readily implemented with superconducting qubits or single electron spins, which exhibit resonances precisely in this frequency range, transduction to optical frequencies requires novel strategies and methodologies. As one key requirement to implement SAW-driven quantum gates with optically active nanosystems, e.g. based on Landau-Zener transitions in a coupled quantum dot-nanocavity system [49,50], the SAW has to be phase-locked to a train of laser pulses. Such quantum gates typically require resonant excitation with highly coherent pulses emitted by a mode-locked laser system, whose pulse repetition rate is fixed. Thus, stable phase-locking between the SAW, whose frequency is set by the IDT's periodicity, and these laser pulses is a challenging task.

Here we report on the application of a wide passband IDT architecture to establish a stable phase-lock between a SAW and a train of laser pulses for advanced acousto-optic spectroscopy of single QDs. Our IDT design is based on a multi-harmonic split-finger architecture [9] for the excitation of a fundamental SAW and a well-defined number of overtones. It is combined with a frequency chirp [51] by introducing a variation of the IDT's periodicity $p$. This combination results in wide frequency bands for each harmonic, which we characterize optomechanically using single sensor QDs. Most importantly, we show that SAWs can be generated by the frequency filtered electrical monitor signal of a free-running mode-locked laser, giving rise to natively stable phase-locking between the train of laser pulses and the generated SAW. We introduce both, electrical and optical delays and tune the relative phase for stroboscopy. Finally, we realize full-fledge SAW-spectroscopy of a single QD. In this scheme, we combine phase-locked excitation with high spectral and high temporal resolution and monitor the dynamic spectral modulation of the QD emission line on sub-nanosecond timescales.

Experimental Implementation

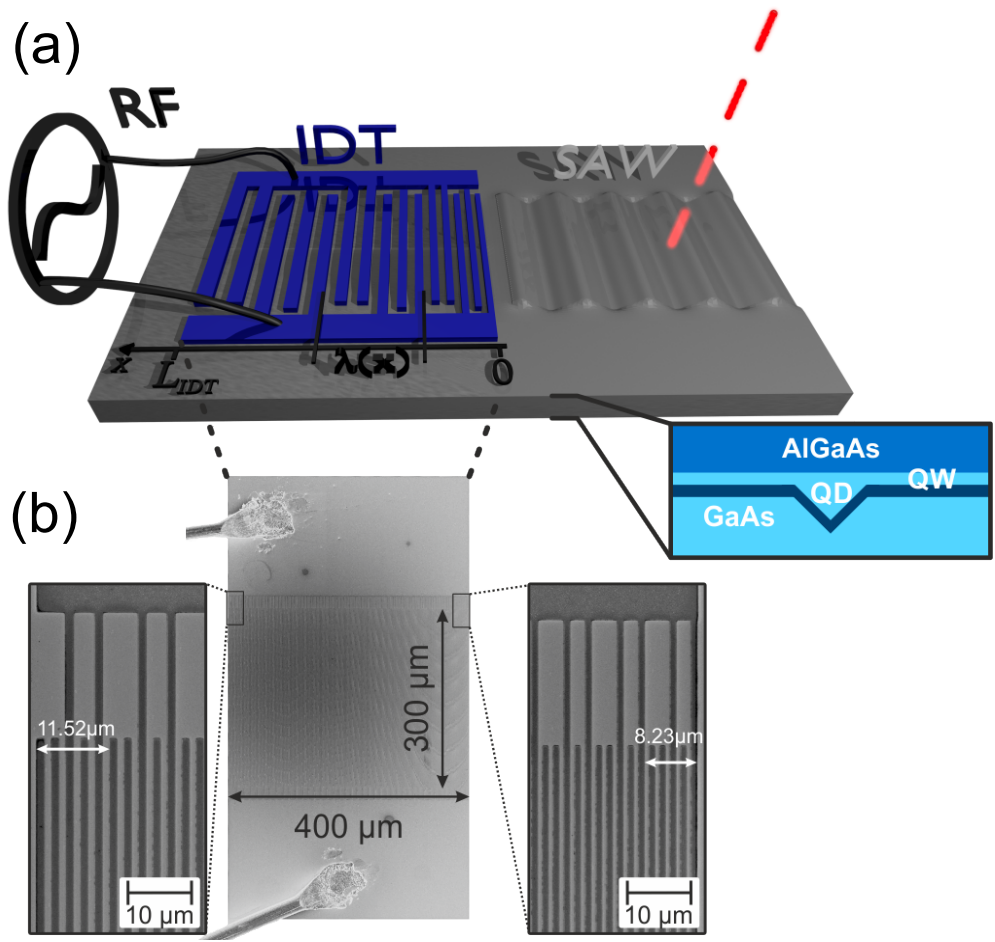

Figure 1 – Device layout and optomechanical sensing using a sensor QD – (a) Left: Device layout consisting of a metal IDT patterned onto an (Al)GaAs based heterostructure. The IDT design combines a Split-52 electrode configuration for multi-harmonics generation and a variation of the periodicity $p$ of the fingers of the IDT. The latter gives rise to a variation of the wavelength of the generated SAW along the axis x of the IDT ($2p = \lambda(x)$). The x direction corresponds to the [110] crystal direction of the (Al)GaAs heterostructure. (Right) The heterostructure contains a single layer of sensor QDs. These are A schematic of our device is depicted in Figure 1 (a). It comprises of IDTs patterned directly on a (Al)GaAs heterostructure containing a single layer of strain-free GaAs/AlGaAs quantum dots (QDs) [52]. Details on the heterostructure can be found in Appendix A. The design of the IDTs is based on a Split-52 configuration, which allows for the excitation of a fundamental SAW and three overtones harmonics $f_{SAW,n} = n \cdot f_{SAW,1}$, with $f_{SAW,1} = 250$ MHz and $n = 1,2,3,4$ [9]. We combined this design with a frequency chirp [53] by introducing a linear variation of the fundamental periodicity

$$2p(x) = \lambda_1(x) = \lambda_0 + \alpha x$$

*Equation 1*

along the IDT axis. In this equation, $\alpha$ is the dimensionless chirp parameter and $0 \leq x \leq L_{IDT}$, with $L_{IDT}$ being the length of the IDT structure. Thus, in the frequency domain the IDT's fundamental resonance and higher harmonics become:

$$f_{SAW,n}(x) = n \cdot c_{SAW}/\lambda_1(x) = n \cdot c_{SAW}/(\lambda_0 + \alpha x).$$

*Equation 2*

$c_{SAW} \approx 2920$ m/s is the phase velocity of the SAW along the [110] direction of the GaAs (001) surface. Moreover, the introduction of a chirp broadens the range of frequencies over which the IDT transduces power between the electrical and acoustic domains to $\Delta f_{SAW,n} = \frac{c_{SAW}}{\lambda_0} - \frac{c_{SAW}}{\lambda_0 + \alpha L_{IDT}}$. For our experiments, we used two different IDTs, IDT1 and IDT2 with $\Delta f_{SAW,1} = 50$ MHz and $\Delta f_{SAW,1} = 100$ MHz, respectively. The aperture of both IDTs, i.e. the length of the overlapping IDT fingers, was $A_{IDT} = 300$ μm, see Fig. 1(b). The full sets of design parameters for both IDTs are summarized in Table 1. The variation of the IDT period can be nicely seen from the scanning electron micrograph recorded at both ends of IDT2, which are presented in Figure 1 (b).

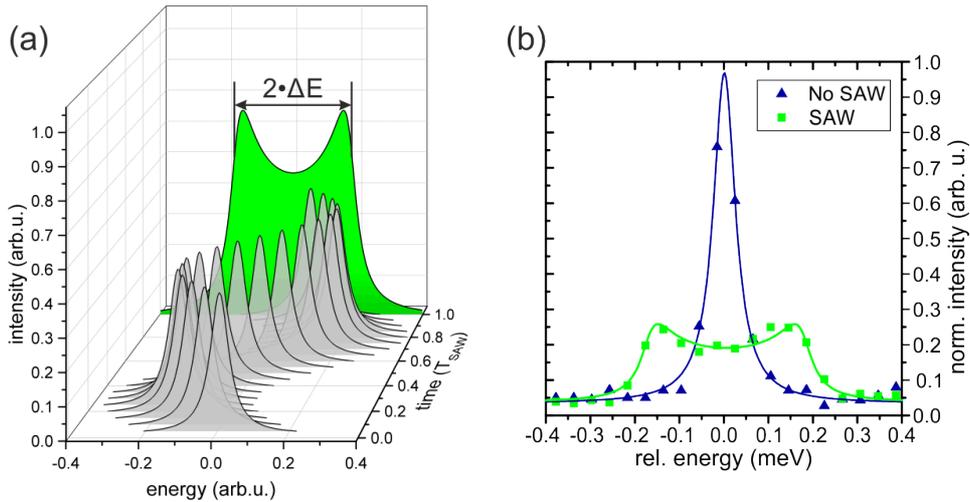

*Figure 2 – Nanoscale pressure sensing – (a) Expected dynamic sinusoidal spectral tuning of a Lorentzian emission line (grey) over one acoustic cycle induced by the SAW's strain field. Integration over all times during the full cycle results in the green averaged spectrum. From fitting Equation 3 results the amplitude of the dynamic broadening, $\Delta E$, can be extracted. (b) Emission of a single QD without a SAW applied (blue symbols) and strained by an 801 MHz SAW (green symbols). Solid lines are best fits of Equation 3 to the experimental data.*

The embedded QDs act as nanoscale pressure sensors and detect the local hydrostatic pressure of a SAW via the deformation potential coupling [45]. These QDs are dynamically strained by the SAW, which gives rise to a spectral broadening of the emission line. The employed (Al)GaAs/GaAs are particularly suited for pressure sensing, because emission lines can be observed over the full acoustic cycle [9]. This is in stark contrast to the more established self-assembled In(Ga)As QDs nucleating on a wetting layer, which forms a two-dimensional transport channel [42,54]. The QDs studied here are almost perfectly decoupled from a continuum of states, in which spatiotemporal carrier dynamics could be induced by the SAW. Thus, carriers are expected to be captured on a 10's of picosecond timescale after photoexcitation [55,56] and reprogramming of the QD's charge state by the SAW is inhibited. Figure 2 (a) illustrates the underlying physics: the Lorentzian emission line (grey) of the QD is is spectrally tuned by the SAW's strain field as time evolves over on SAW-cycle. Averaging of the full acoustic cycle, this dynamic sinusoidal spectral modulation gives rise to a broadening of the emission line. In a time-integrated experiment averaging over $T_{SAW} = 1/f_{rf}$, the broadening is given by

$$I(E) = I_0 + f_{rf} \frac{2A}{\pi} \int_0^{1/f_{rf}} \frac{w}{4 \cdot \left(E - \left(\Delta E \cdot \sin(2\pi \cdot f_{rf} \cdot t)\right)\right)^2 + w^2} dt.$$

*Equation 3*

In this expression, $E$ denotes the energy relative to the center of the unperturbed line $E_0$, $A$ the integrated intensity in absence of SAW, $f_{rf}$ the frequency of the applied electrical voltage signal, and $\Delta E$ the spectral broadening induced by the SAW-modulated deformation potential coupling. The resulting broadened emission line is plotted in green in Figure 2 (a). From this analysis, we evaluate the spectral modulation amplitude $\Delta E(f_{rf})$ and, thus, quantify the local hydrostatic pressure at the position of the QD.

In Figure 2 (b), we demonstrate nanoscale sensing of the SAW's acoustic field. We compare measured time-integrated emission spectra (symbols) of a single QD without (blue) and with (green) a SAW applied. The energy axis is referenced to the center energy of the unperturbed emission line. In the latter case, the QD's Lorentzian emission line is spectrally tuned with frequency $f_{SAW} = 801$ MHz by the time-modulated deformation potential coupling.

*Table 1 – IDT design parameters*

|  | $f_{SAW,1}$ | $\Delta f_{SAW,1}$ | $\lambda_0$ | $\alpha$ | $L_{IDT}$ | $A_{IDT}$ |
|---|---|---|---|---|---|---|
| IDT1 | 250 MHz | 50 MHz | 9.6 µm | $4.8 \cdot 10^{-3}$ | 400 µm | 300 µm |
| IDT2 | 250 MHz | 100 MHz | 8.23 µm | $8.225 \cdot 10^{-3}$ | 400 µm | 300 µm |

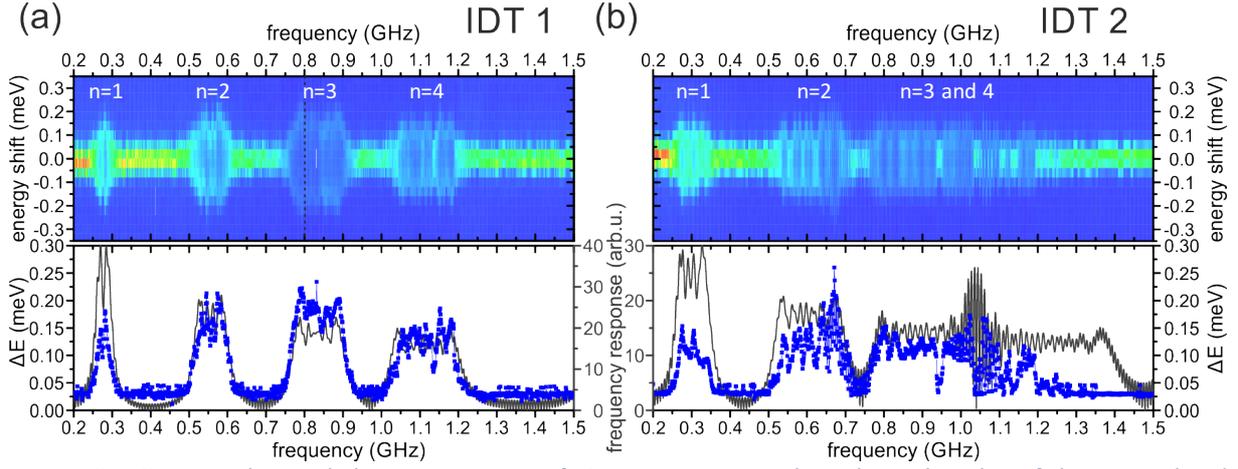

*Figure 3 – Optomechanical characterization of IDTs – Upper panels: False-color plot of the normalized emission intensity as a function of $f_{rf}$ and $\Delta E$ for IDT1 (a) and IDT2 (b) showing clear spectral broadening of the emission line at the characteristic frequency bands $n = 1,2,3,4$ of each IDT. In (a) the dashed vertical line marks the spectrum of the modulated QD shown in Figure 1 (c). Lower panels: $\Delta E$ (blue symbols) derived from the data in the upper panels and calculated frequency response (grey solid line) of IDT1 (a) and IDT2 (b).*

We characterized the SAW generation of both IDTs by measuring the emission spectrum of a single QD as we sweep the radio frequency $f_{rf}$ of the output voltage of a *rf* signal generator, while keeping the *rf* power level constant at $P_{rf} = +24$ dBm (~250 mW). The upper panels of Figure 3 (a) and (b) show results of such frequency sweeps for IDT1 and IDT2, respectively. The normalized emission intensity is color-coded with red (blue) colors corresponding to high (low) intensity and plotted as a function of $f_{rf}$ and photon energy shift relative to the unperturbed QD emission energy $E_0$. $E_0$ has been corrected for a small shift of the center energy due to a variation of the thermal load as $f_{rf}$ is tuned, which was measured independently at times the SAWs has passed by the QD's position [57]. For both IDTs, we observed a pronounced spectral broadening of the QD emission line over broad ranges of $f_{rf}$ for all harmonics expected for our Split-52 configuration. The observed broadening is larger for IDT2 compared to that for IDT1, which directly reflects the larger value of the chirp parameter $\alpha$ (see table 1) and the resulting $\Delta f_{SAW,n}$. Furthermore, the linear dependence $\Delta f_{SAW,n} \propto f_{SAW,n} \propto n$, is nicely resolved for the fundamental SAW and its overtones for each IDT. For IDT2, the $n = 3$ and $n = 4$ frequency bands overlap, giving rise to a quasi-continuous band, over which SAWs can be generated. In such an experiment, we are able to directly assess the frequency-dependent SAW conversion efficiency of an IDT by fitting the experimentally measured line shape using Equation 3 [58]. In the lower panels of Figure 3 (a) and (b), we plot the extracted spectral broadening $\Delta E(f_{rf})$ of the sensor QD emission line (blue symbols) and the calculated frequency response of the two IDTs (black line) as a function of $f_{rf}$. To model the frequency response of the IDT, we used an impulse model [59,60], detailed in Appendix B. Clearly, the optomechanically measured frequency response nicely agrees with our simulation for IDT1. For IDT2, the isolated frequency bands $n = 1$ and $n = 2$ show good agreement between experiment and simulation. For $n = 3$ and $n = 4$, the experimental data exhibits a complex interference pattern. For $f_{rf} \gtrsim 920$ MHz it differs from that expected from our basic impulse IDT model.

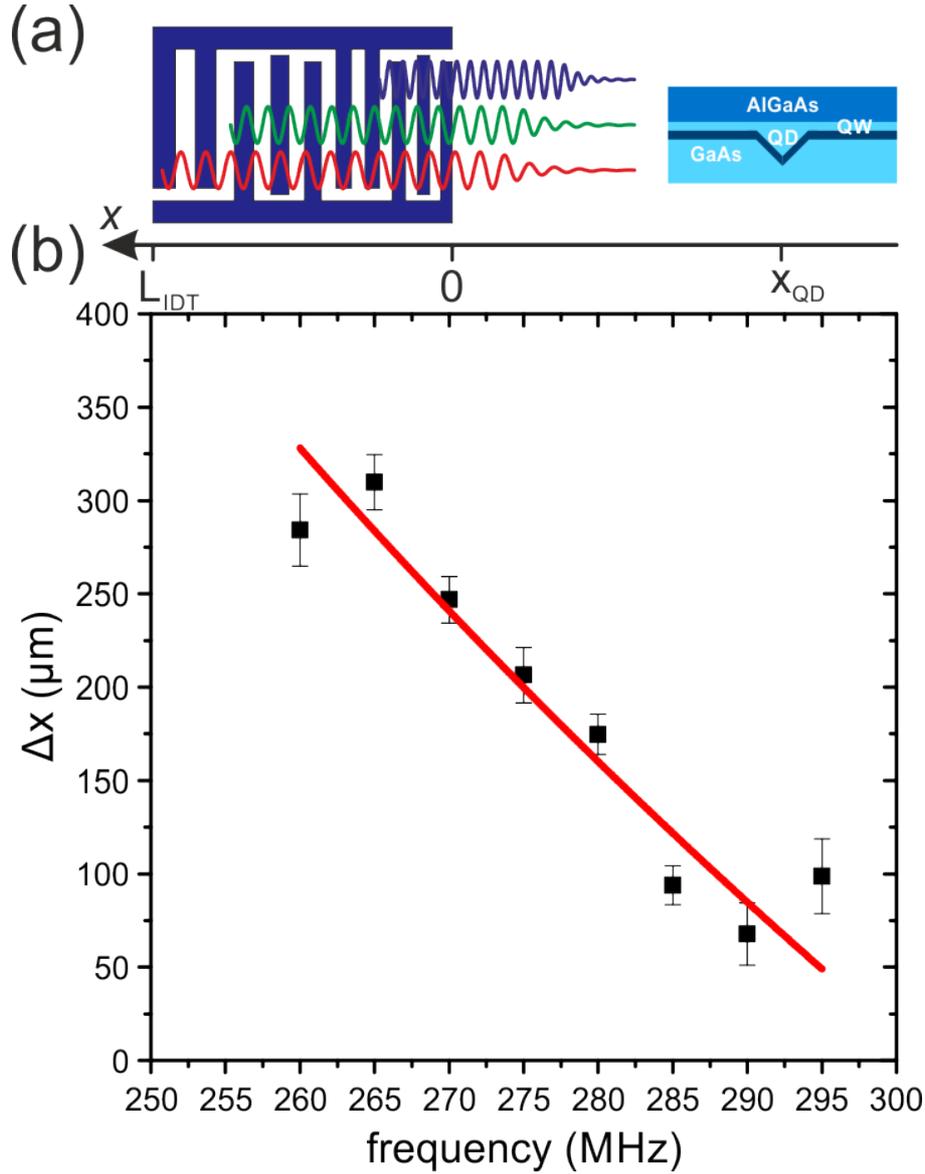

*Figure 4 – Local SAW generation in a chirped transducer – (a) Illustration of the frequency dependent variation of the SAW's propagation length $\Delta x$ from the point of excitation along the axis of the IDT (left) to the fixed position of the sensor QD. (b) Measured $\Delta x$ as a function of frequency across the $f_{SAW,1}$ band of IDT1 (symbols) and best fit of Equation 4 to the data.*

In our chirped transducer design, different $f_{SAW,n}(x)$ are generated at well-defined positions $x$ along the IDT axis. This effect is shown schematically in Figure 4 (a). For our design, the corresponding propagation length scales with the SAW frequency as

$$\Delta x = \frac{n \cdot c_{SAW}}{\alpha} \cdot \frac{1}{f_{SAW,n}} - \frac{\lambda_0}{\alpha}.$$

*Equation 4*

Figure 4 (b) shows the measured $\Delta x$ as a function of $f_{SAW,1}$ of the fundamental resonance band of IDT1. The measured data (symbols) confirms the expected dependence for our design. The red curve is a best fit of Equation 4 to the data. From this fit, we derive $\alpha = (4.9 \pm 0.6) \cdot 10^{-3}$. This value agrees well within the experimental error with the nominal design value of $4.8 \cdot 10^{-3}$ [cf. Table 1].

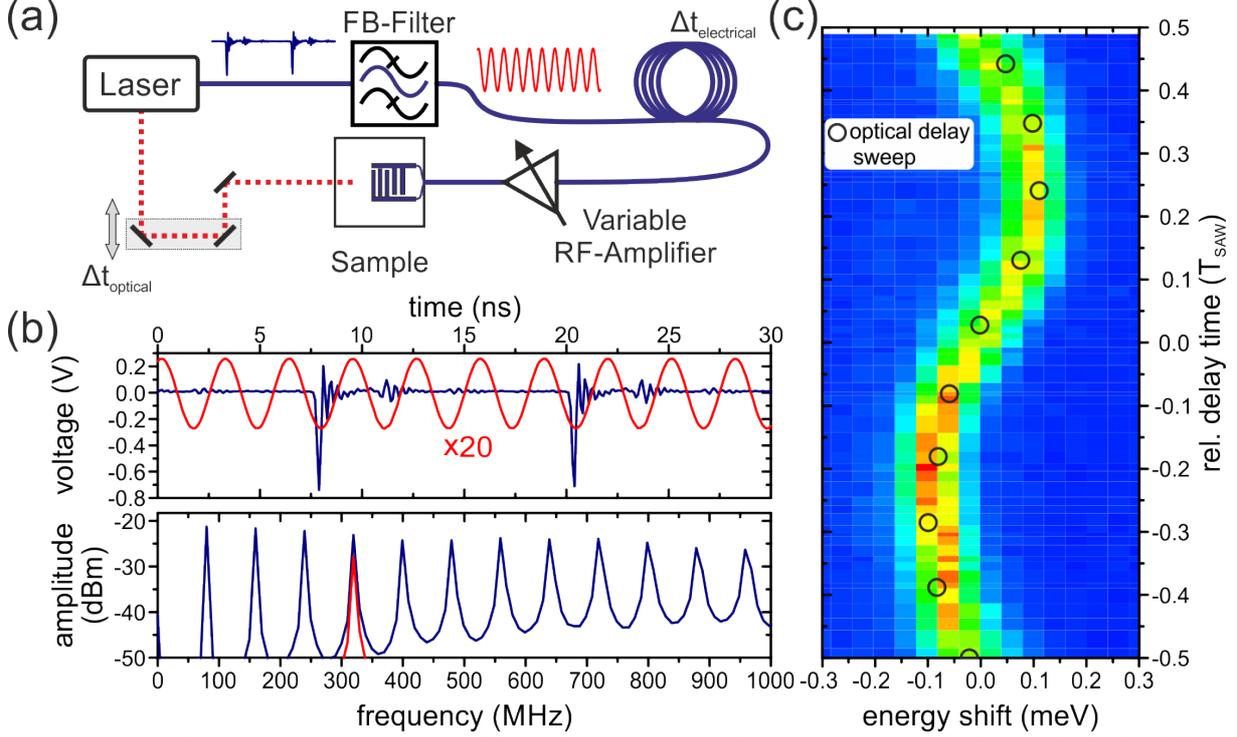

*Figure 5 – Stroboscopic spectroscopy – (a) Implementation: The monitor output of a mode-locked fiber laser passed through a frequency band filter (FB-Filter) to isolate $f_{rf} = 4 \cdot f_{laser} = 320 \, MHz$. The electrical signal is attenuated to a constant power level, amplified to $P_{rf} = +24 \, dBm$ and finally applied to IDT2. The temporal delay between the excited SAW and the train of laser pulses is tuned either electrically by a cable-based delay ($\Delta t_{electrical}$) or optically by variation of the propagation length of the laser beam ($\Delta t_{optical}$). (b) Upper panel: Electrical signal directly from of the laser monitor output (blue) and frequency filtered (red) after passing through the FB-filter. Lower panel: FFT amplitude of the monitor signal (blue) consisting of $f_{laser}$ and 11 overtones and the filtered signal (red) containing only a single component $4 \cdot f_{laser}$. (c) Time-integrated emission spectra recorded from a single sensor QD as a function of $\Delta t_{electrical}$ (at fixed $\Delta t_{optical}$) plotted in false-color representation. The independently measured $\Delta E(\Delta t_{optical})$ for fixed $\Delta t_{electrical}$ is overlaid as open symbols. The characteristic sinusoidal modulation induced by the dynamic acoustic field of the SAW is faithfully reproduced in both independent and complementary experiments.*

We employed the demonstrated *broadband* excitation of SAWs to establish a stable and programmable phase relation between the sound wave and the train of pulses of a mode-locked laser. To ensure a stable phase-lock the laser repetition rate, $f_{laser}$, and $f_{SAW}$ have to be harmonics, $f_{SAW} = m \cdot f_{laser}$, $m$ integer. To meet this strict criterion, the resonance of the transducer has to be precisely matched to $m \cdot f_{laser}$, which is difficult to meet for the narrow band response of a standard IDT without chirp. For the chirped design employed here this shortcoming can be elegantly overcome taking advantage of the tailorable emission band. A schematic of our setup is presented in Figure 5 (a). The mode-locked laser used in our experiments provides the electrical signal shown in blue in the upper panel of Figure 5 (b). In the Fourier analysis of this signal depicted in the lower panel, we identify the repetition rate $f_{laser} = 80 \, MHz$ of the laser pulses. Moreover, 11 higher Fourier components are detected with comparable amplitude. For stroboscopic spectroscopy, we used a frequency band filter to isolate the 3$^{rd}$ overtone at $4 \cdot f_{laser} = 320 \, MHz$, matching the $f_{SAW,2}$ frequency band of IDT2. The filtered signal and its Fourier transform are plotted in red in Figure 5 (b). We adjusted the power level of the frequency filter signal using a variable attenuator. This signal was amplified to a constant power level of $P_{rf} = +24 \, dBm$, directly applied to IDT2 to excite a SAW. Thus, the relative phase (i.e. temporal delay) between the SAW and the train of laser pulses is stably locked. Next, we demonstrated tuning of this relative phase by either introducing a temporal delay in the electrical $\Delta t_{electrical}$ or optical branch $\Delta t_{optical}$ of the setup. As shown in Figure 5 (a), $\Delta t_{electrical}$ was controlled by introducing cable-based electrical delays. The cable-length-dependent losses in this scheme require the two-stage power level control using a variable attenuator and a constant gain amplifier. Two typical stroboscopic scans recorded from a single sensor QD are presented in Figure 5 (c). For the $\Delta t_{electrical}$-scan, the emission intensity is color-coded plotted as a function of spectral shift relative to the unperturbed energy of the emitted photon and the programmed temporal delay over a full acoustic cycle. In these data, we resolve clearly the expected dynamic modulation of the QD emission energy by the oscillating acoustic field of the SAW [9,45]. As mentioned before, no signatures of dynamic charge state programming are resolved in the data. Furthermore, we independently confirmed a tunable temporal delay in the optical branch $\Delta t_{optical}$ by varying the optical propagation length of the laser pulses while keeping $\Delta t_{electrical}$ constant [61]. From the obtained data, we extracted the QD emission energies as function of $\Delta t_{optical}$. Such measured QD emission energies for ten values of $\Delta t_{optical}$, equally distributed over one acoustic cycle, are plotted as symbols in Figure 5 (c). These independently measured QD emission energies, programmed by tuning $\Delta t_{optical}$, faithfully follow the oscillation measured by variation of $\Delta t_{electrical}$. These combined experiments nicely confirm the reciprocity of our two independent approaches to tune the relative phase between the SAW and the laser pulses.

## Phase-locked excitation and time-resolved detection

Finally, we demonstrate full time-domain spectroscopy by combining phase-locked optical excitation and time-resolved detection of the QD emission in the same experiment. The implemented experimental scheme is shown schematically in Figure 6 (a). The laser reference signal is sent to a –3dB (50:50) rf splitter. One port of this splitter is connected to IDT2 to generate a SAW at $4 \cdot f_{laser} = 320$ MHz. Again, the temporal delay between the SAW and the laser pulses exciting the QD (red dashed line) is tuned electrically. The emission of the QD (green dashed line) is spectrally filtered by a monochromator and detected by a single photon avalanche detector (SPAD) [42]. The second port of the rf splitter and the electrical output of the SPAD are directly connected to the inputs of a time-correlated single photon counting (TCSPC) module. The expected temporal evolution in shown schematically on the right of Figure 6 (a). Here, we assumed a sinusoidally modulated emission line. The system is optically excited at $\approx 0.4 T_{SAW}$ and decays exponentially with a time constant of $\tau \approx 0.5 \cdot T_{SAW}$. The signal intensity is color coded and plotted as a function of time (vertical axis) and photon energy (horizontal axis). This model predicts, that in such experiment the time-evolution of the QD emission is directly resolvable. Figure 6 (b-e) demonstrates phase-locked excitation and time- and spectrally resolved detection for four different times of photoexcitation, $\Delta t_{electrical}$, distributed over one acoustic cycle (3.125 ns). The TCSPC counts are normalized for each $\Delta t_{electrical}$ and plotted color-coded as a function of energy shift and absolute time during the acoustic cycle. In contrast to the modelled emission behaviour shown in Figure 6 (a), here the PL decay time of the QD $\tau_{PL} \approx 0.32$ ns $\approx 0.1 \cdot T_{SAW}$. Thus, the expected temporal modulation is only resolved in a relatively short time interval after photoexcitation. The temporal modulation of the QD emission line, derived from our stroboscopy data, is plotted as a solid dashed line. Clearly, all four spectral and temporal evolutions of the emission signal precisely follow the stroboscopy data (dashed line): in Figure 6 (b), photoexcitation occurs at $\Delta t_{electrical} \approx 0.4 \cdot T_{SAW}$, slightly before the maximum of the spectral modulation. Under this excitation condition, no pronounced spectral shift of the emission is observed because the emission line remains approximately fixed in the spectral domain within the decay time of the QD. This changes dramatically, as $\Delta t_{electrical}$ is increased by $\approx 0.25 \cdot T_{SAW}$ in Figure 6 (c). Under these excitation conditions, the emission line shifts rapidly towards higher photon energies. The shift rate, is sufficiently large and leads to the pronounced spectral shift within $\tau_{PL}$ as observed in our data. The direction of this shift changes in Figure 6 (d). Here, the QD is excited on the falling slope of the modulation, again in excellent agreement with the experimental observation. Finally, Figure 6 (e) presents data for $\Delta t_{electrical} \approx 0.05 \cdot T_{SAW}$ set almost exactly to the maximum of the spectral modulation. Here, the emission line initially remains constant and starts to shift to lower photon energy as the shift rate starts to increase. These observations prove that the emission energy is shifted on time-scales shorter than the lifetime of the probed emitters. We note, that for $\tau_{PL} \geq 0.5 \cdot T_{SAW}$, time-resolved detection is imperative: in a simple stroboscopic experiment the observed time-integrated emission converges to that of the phase-averaged case demonstrated in Figure 2.

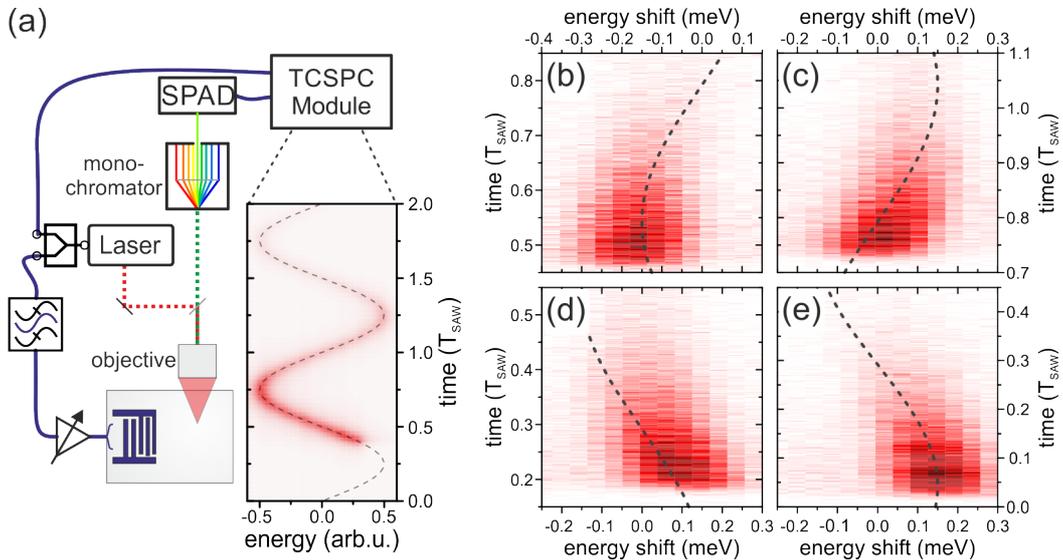

Figure 6 – Phase locked excitation and time-resolved detection – (a) Implementation: The monitor output of a mode-locked fiber laser is split by a -3dB rf splitter. One output is passed through a frequency band filter (FB-Filter) to isolate $f_{rf} = 4 \cdot f_{laser} = 320\ MHz$, attenuated, amplified and connected to IDT2. The emission of a sensor QD is detected by a single photon avalanche photodiode (SPAD). The detector signal is time-correlated with the laser monitor signal from the second output of the rf splitter using a time correlated single photon counting (TCSPC) module. The predicted time- and spectrally resolved emission spectra of a QD is shown in falsecolor representation on the right. (b-e) Time-resolved emission spectra of a single sensor QD recorded for four independently set values of $\Delta t_{electrical}$. The SPAD counts are color coded and plotted as a function of $\Delta E$ and time t during the acoustic cycle. The detected spectral evolution faithfully follows the dynamic modulation by the SAW and the modulation observed by time-integrated stroboscopy (dotted line) shown in Figure 5.

Conclusions and perspectives

In conclusion, we have demonstrated a versatile scheme to realize stable phase-locking between the train of optical pulses emitted by a mode-locked laser and a radio frequency SAW. Our scheme relies on properly designed transducers, which combine a multi-harmonic architecture with a frequency chirp. This combination enables the excitation of SAWs over wide frequency bands for the fundamental SAW and its higher harmonics. Our demonstrated scheme will be the foundation to advanced SAW-based spectroscopy techniques, nanomechanical manipulation and ultimately in acoustic quantum technologies. The frequency chirp is particularly powerful. Its wide passband enables a stable phase lock between the acoustic signal and the coherent optical pulses emitted with a fixed repetition rate by a mode-locked laser. For conventional, narrowband IDTs, even small deviations in the transducer fabrication detunes $f_{SAW} \neq m \cdot f_{laser}$, such that no stable phase can be established. Apart from QDs employed here and other types of quantum emitters [62], the demonstrated method is directly applicable both for acousto-optoelectronic spectroscopy to probe charge carrier dynamics and mobilities [63–65], or phase-locked SAW-photoconductance spectroscopy and tomography [20,66–68]. It will prove particularly useful for any system relying on coherent and resonant optical excitation and probing its acoustically imprinted dynamics on ultrafast timescales, including SAW-driven Landau-Zener gates [49] or exciton-photon polariton condensates in microcavities [69,70]. Moreover, in fused LiNbO$_3$-photodiode hybrids [71], our concept will allow unified electrical and acoustic control ultimately on sub-nanosecond timescales [72–74]. In the emerging field of optomechanics, the recently demonstrated mutually coherent control of optomechanical resonators [34] by coherent SAWs and optical fields could be extended to pulsed optical excitation schemes. Moreover, our chirp design overcomes a native limitation of non-chirped multi-harmonic transducers: the mass loading by the metal electrodes of the transducer leads to a frequency-dependent shift of SAW phase velocity underneath the transducer and thus to a change of the resonance frequency. So, for higher harmonics the relation $f_{SAW,n} = n \cdot f_{SAW,1}$ is no longer valid, rendering multi-component spanning additive Fourier synthesis [9] extremely challenging. This shortcoming is no longer present for the IDT design presented here, paving the way to arbitrary waveform synthesis over the entire hypersonic frequency domain. Such tailored phonon fields enable coherent control schemes in SAW-resonator based quantum acoustics [48] or theoretically conceived SAW-driven quantum gates [49].

## Acknowledgements

This work was supported by the Deutsche Forschungsgemeinschaft (DFG) via the Emmy Noether Program (KR3790/2-1) and the Cluster of Excellence "Nanosystems Initiative Munich" (NIM). AR acknowledges support from the Linz Institute of Technology (LIT) and the FWF (P 29603).

the Emission of Single Excitons in Photonic Crystal Cavities., Nat. Commun. **5**, 5786 (2014).

## Appendix A: Sample design

The (Al)GaAs heterostructure used in our experiment contains a single layer of strain-free GaAs QDs embedded in AlGaAs barriers fabricated by a Ga-droplet etching technique [52]. The full details on the heterostructure can be found in [9]. IDTs (Ti 5 nm/ Al 50 nm) were monolithically fabricated directly on the GaAs-based heterostructure in a lift-off process using a resist mask patterned by electron beam lithography.

## Appendix B: IDT design

The IDT design was derived from an impulse model [59,60]. Here, every finger $n$ of the IDT is considered as the source of a plane wave $A_n \exp\left(i2\pi f \frac{x_n}{c_{\text{SAW}}}\right)$. In this expression $x_n$ is the position of the $n$-th finger. $A_n$ is the amplitude of the wave and reflects the polarity of the different fingers of the IDT. The frequency response is then obtained by summation over all fingers $n$ of the IDT $\sum_n A_n \exp\left(i2\pi f \frac{x_n}{c_{\text{SAW}}}\right)$. We employed the same model to derive the chirp-free multi-harmonic design in Reference [9].

## Appendix C: Experimental details

All experiments were performed at $T$ = 4 K in a closed-cycle Helium cryostat equipped with a low-temperature photonic probe station (PPS, attocube, Munich, Germany). Charge carriers were photogenerated by pulsed laser sources focused to a diffraction limited spot by a NA=0.81 low temperature microscope objective (LT-APO/NIR/0.81, attocube, Munich, Germany). Emission from single QDs was collected by the same objective and dispersed in a 0.75 m grating monochromator and detected either by a liquid nitrogen-cooled charge-coupled device (CCD) for multi-channel time-integrated spectrum acquisition or a single photon avalanche photodiode (SPAD) for single-channel, time-resolved detection.

### Radio frequency scans

For experiments, in which $f_{rf}$ was tuned (Figure 3 and Figure 4), a diode laser emitting $\approx 90$ ps long pulses with a wavelength of 660 nm was triggered with a repetition rate $f_{laser} = 80$MHz. The *rf* voltage output by a signal generator was amplified to a constant power level of $P_{rf} = +24$ dBm and connected to the IDT. The driving signal is applied as $\Delta t = 2000$ ns-long pulses to reduce heating of the sample [41]. The electrical pulses triggering the pulsed laser were only active during the time the SAW propagates across the position of the QD [40]. $f_{laser}$ and rate $f_{rf}$ were incommensurate ($f_{rf} \neq m \cdot f_{laser}$, $m$ being integer) to average the dynamic modulation of the QD over a full *rf* cycle in a single time-integrated spectrum [71].

### Phase-locked excitation with tunable laser of fixed repetition rate

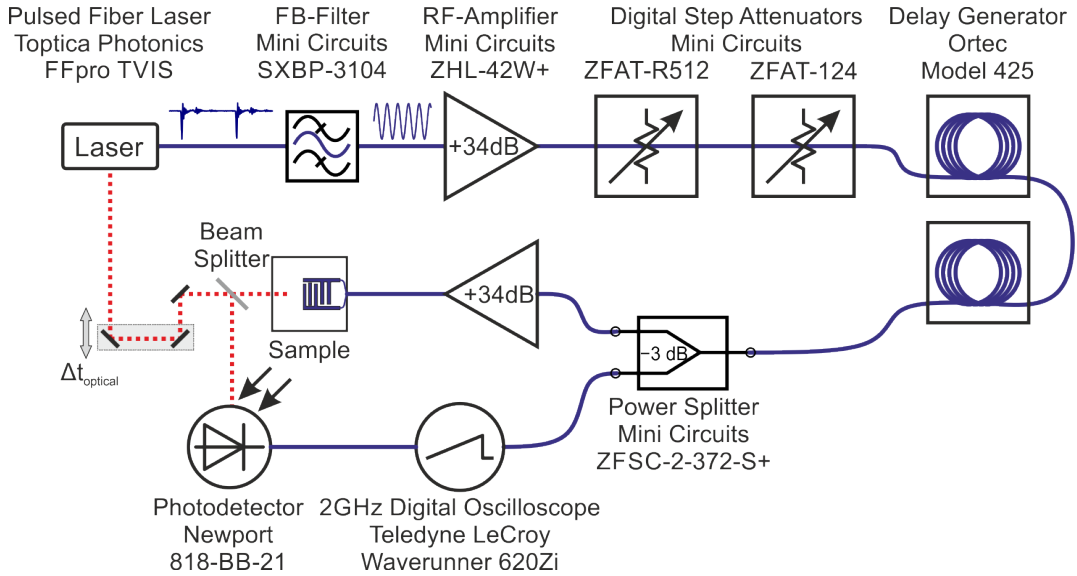

*Figure 7 – Detailed schematic of the setup used for the full phase-locked excitation and time-integrated detection (data shown in Figure 5(c)). The electrical signal is depicted by the blue connections, the optical by the red ones.*

Figure 7 shows the full setup for stroboscopy using a mode-locked laser [c.f. Figure 5]. A mode-locked tunable fiber laser (Toptica, Gräfelfing, Germany) emitting < 1 ps pulses with a fixed repetition rate $f_{laser} = 80$ MHz of wavelength 640 nm was used for optical excitation. SAW generation was realized as described in the main text. In essence, a frequency bandpass filter (Mini Circuits SXBP-310+) isolated the $4 \cdot f_{laser} = 320$ MHz component of the NIM-pulse electrical monitor output of the laser. Tuning of $\Delta t_{electrical}$ was set by a cascade of two cable based electrical delay systems (Ortec, Modell 425). For each set $\Delta t_{electrical}$, the electrical power was adjusted to $P_{rf} = -10$ dBm by using a constant gain amplifier (Mini Circuits, ZFSC-2-372-S+) and two digital step attenuators (Mini Circuits, ZFAT-R512 and ZFAT-124) to compensate for the inevitable cable length dependent losses and to ensure SAW generation with $P_{rf} = +24$ dBm after amplification by +34dB using a second constant gain amplifier (Mini Circuits, ZFSC-2-372-S+). In order to monitor the electrical power, the signal was split by a -3dB rf splitter (Mini Circuits, ZFSC-2-372-S+) before final amplification and analysed by a 2 GHz bandwidth digital storage oscilloscope (Teledyne LeCroy, Waverunner 620Zi). Furthermore, the oscilloscope was used to determine the exact relative phase between the electrically excited SAW and the optical excitation using a high-speed photodetector (Newport, 818-BB-21) to monitor the optical output of the laser. Tuning of $\Delta t_{optical}$ was realized manually, by variation of the length of the optical excitation path. For all stroboscopic data shown in Figure 5 (c), time-integrated multi-channel detection was used.

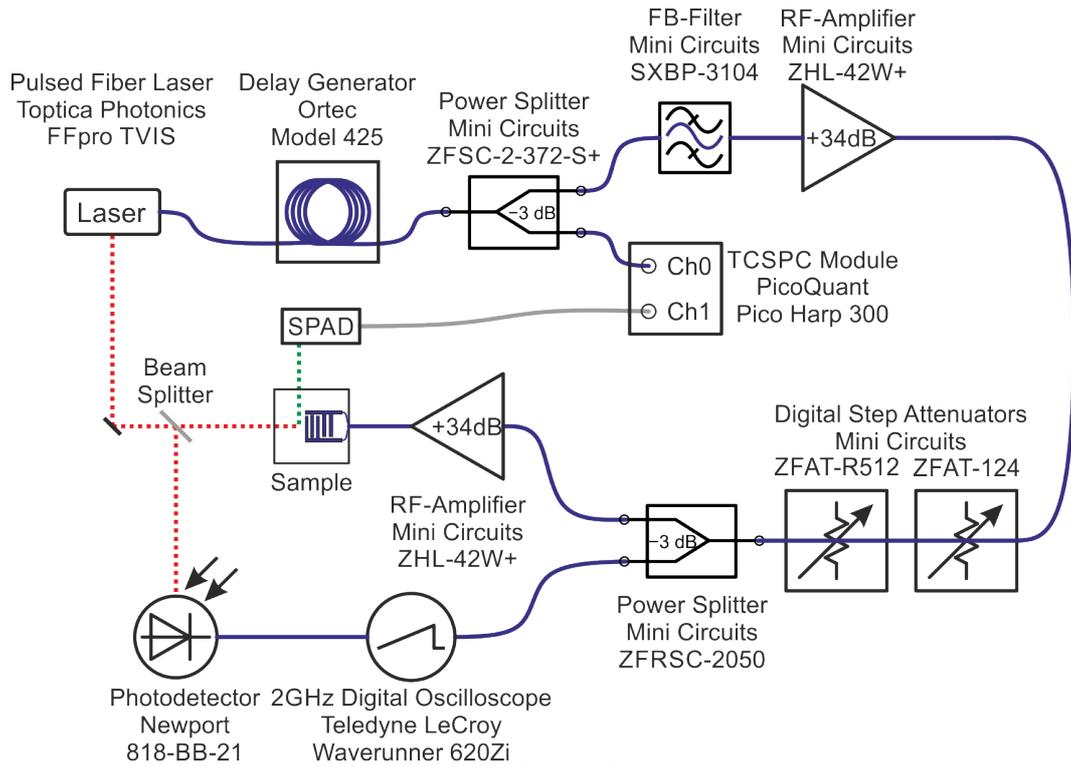

*Figure 8 – Detailed schematic of the setup used for the full phase-locked excitation and time-resolved detection (data shown in Figure 6(b)). The electrical signal is depicted by the blue connections, the optical by the red ones.*

Figure 8 shows a schematic of the full setup for the full phase-locked excitation and time-resolved detection scheme [c.f. Figure 6]. We scanned the detection wavelength of the monochromator and recorded individual PL time transients using a Si-single photon counting module connected to a two-channel TCSPC module (PicoQuant, PicoHarp 300) [32]. In addition, the electrical setup was slightly changed to meet the new requirements. Essentially, one additional -3dB rf splitter (Mini Circuits, ZFSC-2-372-S+) was used to obtain the reference signal for the TCSPC-module. This was done before amplification/attenuation and after setting the electrical delay $\Delta t_{electrical}$, in order to match the input specifications of the TCSPC module and, at the same time, to resolve the set time delay in the time resolved measurements.